\documentclass[pra,showpacs,floatfix]{revtex4}

\usepackage{amsmath}
\usepackage{amssymb}
\usepackage{graphicx}

\begin{document}

\title{Make Slow Fast - how to speed up interacting disordered matter  } 

\author{Goran Gligori\'{c}$^{1,2}$, Kristian Rayanov$^{1}$, Sergej
Flach$^{1,3}$}

\affiliation{$^{1}$Max-Planck-Institut f\"{u}r Physik komplexer Systeme
- N\"{o}thnitzer Strasse 38, D-01187 Dresden, Germany \\
$^{2}$Vin\v{c}a Institute of Nuclear Sciences, University of Belgrade - P.
O. Box 522, 11001 Belgrade, Serbia \\
$^{3}$New Zealand Institute for Advanced Study, Massey University, Auckland,
New Zealand}

\date{\today} \widetext


\begin{abstract}
Anderson and dynamical localization have been experimentally observed with
ultra-cold atomic matter. Feshbach resonances are used to efficiently
control the strength of interactions between atoms. This allows
to study the delocalization effect of interactions for localized wave packets.
The delocalization processes are subdiffusive and slow, thereby limiting
the quantitative experimental and numerical analysis. 
We propose an elegant solution of the problem  by proper ramping the interaction
strength in time. We demonstrate that subdiffusion is speeded up to normal
diffusion
for interacting disordered and kicked atomic systems. The door is open to test
these theoretical results experimentally, and to attack similar computational
quests in higher
space dimensions
\end{abstract}

\maketitle

\subsection*{Introduction}

The quantum wave nature of ultracold atoms in optical potentials
\cite{reviewBloch}, as demonstrated impressively through their
macroscopic condensation \cite{ketterleetc, ketterle}, is the key ingredient
for the recent observation of
Anderson localization with quantum atomic matter \cite{Bi,Ro}. 
Quasi-one-dimensional elongated traps are modulated randomly with speckle
potentials \cite{CHDJWH, CVRSAB}, or simply quasiperiodically with interfering
laser beams \cite{LFFGMWI}, in order to observe the halt of spreading of an
initially localized
wave packet of $10^4$ - $10^5$ Rb and K atoms, and an exponentially localized
atomic density distribution profile.
The length scales are controlled by the localization length $\xi$ which is a
function of the potential parameters, and possibly also the energies
of packet atoms. 
This phenomenon of wave localization is inherently relying on the phase
coherence
of matter waves. It is closely related
to the dynamical localization of the quantum kicked rotor in momentum space,
which was successfully probed 
already in 1995 using ultra-cold Na atoms \cite{MRBSR}. Recent experiments with
quasiperiodically kicked rotors with Cs atoms extend to two- and
three-dimensional disorder potentials \cite{CLGDSG}. Interestingly
systems of one- and two-dimensional optical waveguides have been also recently
used to probe Anderson localization \cite{LAPSMCS, SBFS}.

For some atomic species (K, Cs, Na, Li) Feshbach resonances can be used to
efficiently control the strength of interactions between atoms
\cite{RZECMSIM, WHMNG, IASMSK, KSFBCCCS}.
This opens the possibility to study the fate of Anderson localization for
interacting localized wave packets. Indeed the first experiment of this kind
\cite{LDTRZMLDIM} showed that interaction beats localization, but in a very slow
way - the second moment $m_2$ of an atomic wave packet increases subdiffusively
in time:
$m_2 \sim t^{\alpha}$ with $\alpha < 1$. This process may stop in the long run
once the atomic density $n$ of the wavepacket reaches the inverse of the
localization length $\xi$
(which touches the quantum world of many-body localization \cite{BAA}). On
shorter times (when typically more than 10 atoms occupy one local single
particle state) the
mean field approximation is a reasonable tool for the study of the subdiffusive
process.  

\subsection*{The problem}
The mean-field approximation replaces the many body linear Schr\"odinger
equation in
a hugely dimensional Hilbert space with a nonlinear Schr\"odinger equation
(NLS),
e.g. the Gross-Pitaevsky equation. 
The effective interaction strength $\beta$ is proportional to the scattering
length $a_s$.
What matters in terms of quality here, is the fact that for almost any disorder
(or quasiperiodic) 
potential realization the corresponding NLS will be nonintegrable.
This seemingly unimportant mathematical property has a very profound impact -
the dynamics of a wave packet becomes in general chaotic in time, characterized
by positive
Lyapunov coefficients and exponential divergence of nearby trajectories. As a
consequence the coherence of phases of waves which constitute a given initial
wave packet is lost,
and with it also the whole effect of wave localization. First observed in
1993
by Shepelyansky for an NLS version of the quantum kicked rotor \cite{DS}, it was
recently studied with great detail for random and quasiperiodic potentials
\cite{LBKSF, MAP, LDM, GS, GBF}. The main outcome for quasi-one-dimensional
models
with two-body atomic interactions is an asymptotic wavepacket spreading with
$\alpha=1/3$ \cite{FKS}. With inverse time units equal to single particle
kinetic energies 
the crossover from 
intermediate to universal asymptotic dynamics takes place at dimensionless time
$\tau \sim 10^6$. In the Florence setup \cite{LDTRZMLDIM} largest times reached
are 
$\tau \sim 10^4$, leaving the experiments in the intermediate
case-to-case-dependent dynamics. Although the onset of subdiffusion is clearly
observed, no reliable experimental
data are currently at hand to measure the exponent $\alpha$, as follows
from the data analaysis and the large statistical errors in \cite{LDTRZMLDIM}.
While some experimental optimization and increase of the kinetic energy may add
one order of magnitude in time, another one-two orders are needed and are
probably currently out of
range of accessibility.
Notably similar problems of insufficient available time scales arise with
computational studies when turning to higher dimensional analogs \cite{GS, LBF, 
MP}.
While two-dimensional models appear to be at the edge of reasonable analysis,
three-dimensional are clearly not.
As follows from the above we can diagnose the problem of lacking time scales for
a safe observation and study of subdiffusive interacting atomic cloud dynamics
in disordered media.

\subsection*{The solution}
Instead of trying to substantially increase available time scales, we propose
here to speed up the subdiffusive process itself. 
This is done by a temporal ramping of the two-body interaction strength,
which can be varied e.g. for K atoms by three orders of magnitude close to the
Feshbach resonance \cite{RZECMSIM}. Why should that help? The momentary
diffusion rate $D$ of
a spreading packet in one spatial dimension is proportional to the fourth power
of the product of interaction strength $\beta$ and particle density $n$: $D \sim
(\beta
n)^4$ \cite{SF}. In the course of
cloud spreading the density $n$ decreases, and therefore also $D$. This is the
reason for the predicted subdiffusion process, which is substantially slower
than normal diffusion.
We propose here to compensate the decrease of the density $n$ with an increase
in the interaction strength $\beta$. Depending on the concrete time dependence
$\beta(\tau)$
we expect different faster subdiffusion processes, and possibly even normal
diffusion. The condition for that outcome to be realized is, that the internal
chaos time scales
(basically the inverse Lyapunov coefficients) will be still short enough so that
the atomic cloud can first get chaotic, and then spread. 
With that achieved, the cloud spreading will be faster, and we can expect that
the available experimental time will suffice for the precise observation
and analysis of the process.

Let us get into numbers for one spatial dimension. The second moment is $m_2
\sim 1/n^2$ and the momentary diffusion constant $D \sim (\beta n)^4$. For a
constant $\beta$ the 
solution of $m_2 =D\tau$ yields $m_2\sim 1/n^2 \sim \tau^{1/3}$, and therefore
$n\sim \tau^{-1/6}$. Thus we choose now a time dependence $\beta \sim
\tau^{\nu}$.
Then the resulting spreading is characterized by 
\begin{equation}
m_2 \sim \tau^{(1+4\nu)/3}\;, \;d=1\;. \label{result1d}
\end{equation}
For $\nu=1/2$ we already obtain normal diffusion $m_2 \sim \tau$.

Similar for two spatial dimensions, where $m_2 \sim 1/n$, for a constant
$\beta$ the cloud spreading is even slower with $m_2 \sim \tau^{1/5}$. With a
time dependent ramping
$\beta \sim \tau^{\nu}$ the resulting speedup is 
\begin{equation}
m_2 \sim \tau^{(1+4\nu)/5}\;,\; d=2\;. \label{result2d}
\end{equation}
For $\nu=1$ we again obtain normal diffusion.

Once ramping is too fast, we
expect to see several different scenaria. Either fragmenting atomic
clouds appear since some parts of the cloud get self-trapped \cite{KKFA, VF} and
some other parts do not.
If self-trapping is avoided, we may also see ramping-induced diffusion: while
the
internal cloud dynamics does not suffice to decohere phases, initial
fluctuations in the
density distribution can lead to considerably different temporal energy
renormalizations in
different cloud spots, and therefore to an effective dephasing similar to a
random noise process
in real time and space.  

\subsection*{Results in one dimension}
Here we study the spreading of atomic clouds in one-dimensional disorder
potentials and in a quantum kicked rotor with interacting atoms. The first model
is described with the discrete NLS (DNLS)
\begin{equation}
 i\frac{\partial{\psi_l}}{\partial \tau}
=\epsilon_l\psi_l+\beta\left(\tau\right)\left|\psi_l\right|^ { 2 }
\psi_l-\psi_{l+1}-\psi_{l+1} \label{dnls}
\end{equation}
in which the on-site energy $\epsilon_l$ is chosen uniformly from a
$\left[-W/2,W/2\right]$ random distribution. The nonlinear quantum kicked rotor
(NQKR) is studied within the diagonal interaction approximation introduced by
Shepelyansky in \cite{DS}:
\begin{equation}  
\psi_{l}(\tau+1)=\sum_{m} (-i)^{l-m} J_{l-m}(k) \psi_{m}(\tau)
e^{-i\frac{\bar{\tau}}{2}m^2
+i\beta(\tau)\left|\psi_{m}\right|^{2}} , \label{eq:NLS_Shep}
\end{equation}
where $\psi_l(\tau)$ are the Fourier coefficients of the corresponding
time-dependent many body
wave function. $J_{l-m}(k)$ is a Bessel function of the first kind, whose
argument $k$ is the kick strength, and $\bar{\tau}$ is
a parameter which relates the period of applied kicks $T$ (set to 
$T=1$) to the natural frequency of rotor, defined as $\omega=\hbar/2M$
($M$-mass of atoms).
In both models $\beta$ is the interaction strength ramped in time $\tau$ -
the dimensionless time for the DNLS and the number of kicks for the
NQKR model:
\begin{equation}
 \beta(\tau) = \left\{
  \begin{array}{l l}
    \beta_0 & \quad ,\tau\leq \tau_0\\
    \beta_0\left(\frac{\tau}{\tau_0}\right)^{\nu} & \quad ,\tau>\tau_0
\label{beta_t}
  \end{array} \right.
\end{equation}
In both models we consider a wave packet which is initially concentrated on a
single site for purely technical reasons, without any loss of generality
(extended initial clouds are perfectly usable as well \cite{SKKF}).
After some first time scale $\tau_0$ the packet spreads approximately over one
localization length $\xi$.
The total norm of the packet is set to one without any loss of generality and is
proportional to the total number of atoms in a cloud 
(similarly, we could also choose any larger norm and rescale
$\beta$ accordingly).  
To characterize the spreading of the cloud we compute
the density $n_l=\left|\psi_{l}(\tau)\right|^2$, the
participation number $P=1/\sum_{l}n^{2}_l$ (the number of strongly excited
sites), and the second moment $m_{2}=\sum_{l}(l-\bar{l})^{2}n_{l}$ (the
squared
distance between the wave packet tails), where
$\bar{l}=\sum_{l}ln_{l}$ is the first moment. In the NQKR the average energy of
the atomic cloud is proportional to the corresponding second moment,
$E=\frac{1}{2}m_{2}$. Another remarkable difference between both systems is that
for large values of $\beta$ self-trapping can occur for 
atomic clouds in disordered spatial systems which may lead to soliton formation
\cite{FW}. For the kicked rotor case this is impossible since
the interaction strength $\beta$ action in (\ref{eq:NLS_Shep}) is cyclic
reflecting the circumstance of periodic kick action in momentum space \cite{FI}.

Equation (\ref{dnls}) was time evolved using a SABA-class symplectic integration
scheme and equation (\ref{eq:NLS_Shep}) as an iteration
map.
The parameters were fixed to $\beta_{0}=1$ and $W=4$ for the DNLS and
$\beta_{0}=0.4$,
$\bar{\tau}=1$ and $k=3$ for the one-dimensional NQKR. Different realizations for the DNLS were
produced by choosing different unique random sequences in the interval
$[-W/2,W/2]$, while for the NQKR they were realized by exciting different
initial states.

\begin{figure}[ht!]
\begin{center}
\includegraphics[scale=0.5]{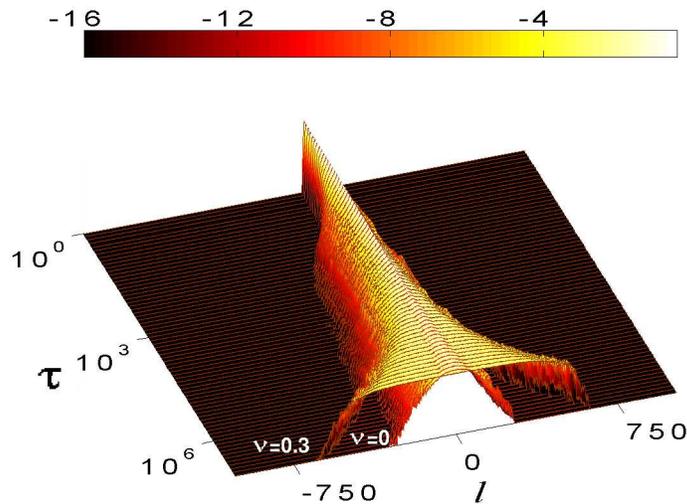}
\caption{Evolution of the averaged norm density $<n_{l}(\tau)>$ in the case
without ($\nu=0$) and with ramping ($\nu=0.3$) in log scale for the DNLS
model.} \label{fig1}
\end{center}
\end{figure}

\begin{figure}[ht!]
\begin{center}
\includegraphics[scale=0.3]{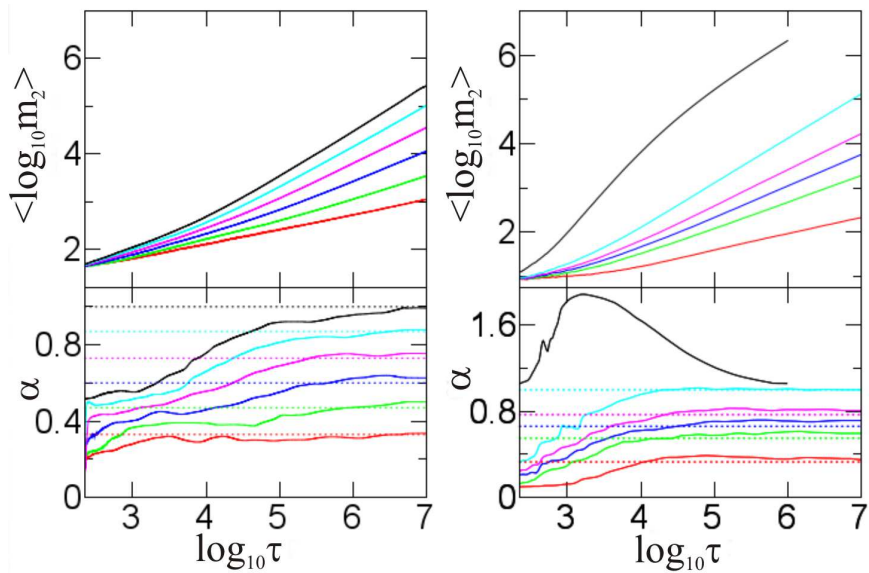}
\caption{\textit{Left column:} the second moments (upper) and their power-law
exponents $\alpha$ (lower) for the DNLS model for $\nu=0$ (red), $\nu=0.1$
(green),
$\nu=0.2$ (blue), $\nu=0.3$ (magenta), $\nu=0.4$ (cyan), and $\nu=0.5$ (black).
\textit{Right column:} the second moments (upper) and their power-law
exponents $\alpha$ (lower) for the NQKR model for $\nu=0$ (red), $\nu=0.17$
(green),
$\nu=0.25$ (blue), $\nu=0.33$ (magenta), $\nu=0.5$ (cyan), and $\nu=1.5$
(black). Dashed colored lines correspond to expected values for exponents in
both cases.} \label{fig2}
\end{center}
\end{figure}

The spreading of wave packets in the DNLS model, without and with ramping of
the nonlinearity are shown in Fig. (\ref{fig1}). Clearly packets 
spread faster when the nonlinearity is ramped in time. To quantify the
spreading exponent, we 
averaged the logs (base 10) of $P$ and $m_2$ over 1000 different
realizations and smoothened additionally with locally weighted regression
\cite{CD}. 
The (time-dependent) spreading exponents are obtained through 
central
finite difference method \cite{Ho}, 
$\alpha=\frac{d<\log_{10}(m_2)>}{d(\log_{10}(\tau))}$.
The results 
for the DNLS and NQKR model are shown in Fig. (\ref{fig2}). The exponents
of
subdiffusive spreading reach the theoretically predicted values. Moreover, for
faster ramping 
of nonlinearity, the asymptotic state with constant exponent  is reached faster.
Monitoring of the participation
number $P$ for the DNLS
indicates that self-trapping starts to occur already for $\nu=0.4$. Results for
the NQKR model, in which the self-trapping is avoided, confirm the reaching of a
normal diffusion process
for $\nu=0.5$. Remarkably, the absence of self-trapping for the NQKR results in
superdiffusion 
on intermediate times for $\nu>0.5$. Finally, the exponent relaxes back to the
normal diffusion value indicating the realization of ramping-induced diffusion.
This case is illustrated for $\nu=1.5$ in Fig. (\ref{fig2}).

\subsection*{Results in two dimensions}

To speed up subdiffusive processes in higher dimensions
we considered the two-dimensional NQKR model based on the map introduced in
\cite{DF}, with an additional phase term which takes into account
interactions in the diagonal approximation:
\begin{eqnarray}  
\psi_{l_{1},l_{2}}(\tau+1)&=&\sum_{s_{1},s_{2}} (-i)^{s_{1}+s_{2}}
J_{s_{1}}(k/2) J_{s_{2}}(k/2) \psi_{l_{1}-s_{1}-s_{2},l_{2}+s_{1}-s_{2}}(\tau)
\nonumber \\
&&e^{-i\frac{\bar{\tau}}{2}((l_{1}-s_{1}-s_{2})^2+(l_{2}+s_{1}-s_{2})^2)
+i\beta(\tau)\left|\psi_{l_{1}-s_{1}-s_{2},l_{2}+s_{1}-s_{2}}(\tau)\right|^{2}}.
\label{eq:2NQKR}
\end{eqnarray}
The notation is the same as in the one-dimensional case, except that now we have
two indices, for each possible direction. Note that according to the relation
for Bessel functions $J_{-n}(x)=(-1)^{n}J_{n}(x)$, the wave packet
$\psi_{l_{1},l_{2}}$, defined by expression
(\ref{eq:2NQKR}), exhibits symmetry with respect to $l_{1}$ and
$l_{2}$ direction. The density is defined as
$n_{l_{1},l_{2}}=\left|\psi_{l_{1},l_{2}}(\tau)\right|^{2}$, the participation
number as $P=1/\sum_{l_{1},l_{2}}n^{2}_{l_{1},l_{2}}$, and the second moment is
$m_2=\sum_{l_{1},l_{2}}[(l_{1}-\bar{l_{1}})^{2}+(l_{2}-\bar{l_{2}})^{2}]n_{l_{1}
,l_{2}}$, where $\bar{l_{1}}=\sum_{l_{1},l_{2}}l_{1}n_{l_{1},l_{2}}$ and 
$\bar{l_{2}}=\sum_{l_{1},l_{2}}l_{2}n_{l_{1},l_{2}}$. Equation (\ref{eq:2NQKR}) was time evolved as an iteration
map for the fixed parameters $\beta_{0}=0.4$, $\bar{\tau}=1$ and $k=2$.

\begin{figure}[ht!]
\begin{center}
\includegraphics[scale=0.5]{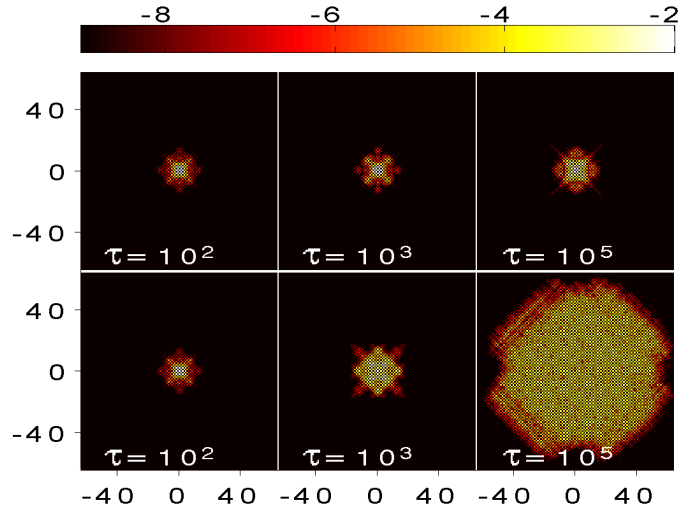}
\caption{The norm denisty $n_{l_{1},l_{2}}(\tau)$ in the case without ($\nu=0$)
(upper row)
and with ramping ($\nu=1.0$) (lower row) after $\tau=10^2$, $\tau=10^3$, and
$\tau=10^5$
kicks in log scale for the two-dimensional NQKR model.}\label{fig3}
\end{center}
\end{figure}

\begin{figure}[ht!]
\begin{center}
\includegraphics[scale=0.35]{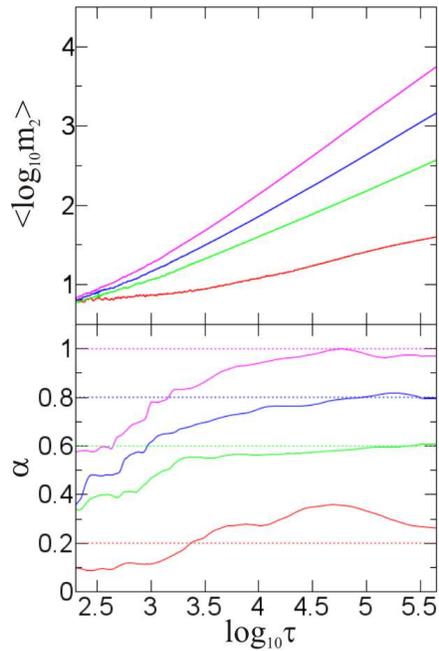}
\caption{The second moments (upper) and their power-law
exponents $\alpha$ (lower) for the two-dimensional NQKR model for $\nu=0$ (red),
$\nu=0.5$
(green), $\nu=0.75$ (blue), and $\nu=1$ (magenta). Dashed colored lines
correspond to
expected values for exponents.} \label{fig4} 
\end{center}
\end{figure}

In Fig. (\ref{fig3}) we compare the wave packet evolution for $\nu=1$
(normal diffusion) and $\nu=0$ at three different moments of time. We clearly
observe the symmetry of wave packet and
a much more violent spreading in the presence of ramping.
The spreading 
exponents are computed similar to the one-dimensional case. We find very good
agreement
with the theoretical prediction (Fig. (\ref{fig4})). Again the asymptotic spreading
state is reached faster for stronger ramping.

\subsection*{Conclusion}

We have investigated the speeding up of the subdiffusive spreading in
interacting disordered and kicked atomic systems by a proper ramping of the
interaction strength in time. We confirm that ramping the interaction strength
leads to faster subdiffusion. For fast enough ramping we even reach normal
diffusion
of atomic clouds.
Self-trapping effects in disordered systems are limiting further speed up of
the 
wave packet spreading. Most importantly the concept works equally well in
one-dimensional
and two-dimensional systems. 

Our results on how to speed up slow subdiffusive processes in
interacting disordered matter will be useful for quantitative experimental and
computational 
studies of the impact of interactions on disorder induced matter wave
localization.
This is particularly true for experimental
realizations with ultra-cold atoms, where the scattering length and thus the
interactions strength can be tuned via Feshbach resonances by magnetic field
variations.
On the other hand, our results are also very useful for computational
studies of spreading regimes in higher dimensional systems, where even modern
computers
reach their limits before reaching the subdiffusive asymptotics.

\acknowledgments G.G. acknowledges support from the Ministry of Education and
Science of Serbia (Project III45010).


\begin{thebibliography}{99}
\bibitem{reviewBloch}{I. Bloch, J. Dalibard, W. Zweger}, Rev. Mod. Phys.
\textbf{80}, 885 (2008)
\bibitem{ketterleetc}{J. R. Anglin, and W. Ketterle}, Nature \textbf{416}, 211
(2002)
\bibitem{ketterle} {W. Ketterle}, Rev. Mod. Phys. \textbf{74}, 1131 (2002)
\bibitem{Bi}{J. Billy, V. Josse, Z. Zuo, A. Bernard, B. Hambrecht, P. Lugan,
D. Cl\'{e}ment, L. Sanchaez-Palencia, P. Bouyer, and A. Aspect}, Nature
\textbf{453}, 891 (2008)
\bibitem{Ro}{G. Roati, C. D'Errico, L. Fallani, M. Fattori, C. Fort,
M. Zaccanti, G.Modugno, M. Modugno, and M. Inguscio}, Nature \textbf{453}, 895
(2008)
\bibitem{CHDJWH}{Y. P. Chen, J. Hitchcock, D. Dries, M. Junker, C. Welford, and
R. G. Hulet}, Phys. Rev. A \textbf{77}, 033632 (2008)
\bibitem{CVRSAB}{D. Cl\'{e}ment, A. F. Varon, J. Retter, L. Sanchez-Palencia, A.
Aspect, and P. Bouyer}, New J. Phys \textbf{8}, 165 (2006)
\bibitem{LFFGMWI}{J. E. Lye, L. Fallani, C. Fort, V. Guarrera, M. Modugno, D. S.
Wiersma, and M. Inguscio}, Phys. Rev. A \textbf{75}, 061603 (2007)
\bibitem{MRBSR}{F. L. Moore, J. C. Robinson, C. F. Bharucha, B. Sundaram, and
M. G. Raizen}, Phys. Rev. Lett. \textbf{75}, 4598 (1995)
\bibitem{CLGDSG}{J. Chab\'{e}, G. Lemair\'{e}, B. Gr\'{e}maud, D. Delande, P.
Szriftgiser, and J. C. Garreau}, Phys. Rev. Lett. \textbf{101}, 255702 (2008)
\bibitem{LAPSMCS}{Y. Lahini, A. Avidan, F. Pozzi, M. Sorel, R. Morandotti, D. N.
Christodoulides, and Y. Silberger}, Phys. Rev. Lett. \textbf{100}, 013906 (2008)
\bibitem{SBFS}{T. Schwartz, G. Bartal, S. Fishman, and M. Segev}, Nature
\textbf{446}, 52 (2007)
\bibitem{RZECMSIM}{G. Roati, C. D'Errico, J. Catani, M. Modugno, A. Simoni, M.
Ignuscio, and G. Modugno}, Phys. Rev. Lett \textbf{99}, 010403 (2007)
\bibitem{WHMNG}{T. Weber, J. Herbig, M. Mark. H. N\"{a}gerl, and R. Grimm},
Science \textbf{299}, 232 (2003)
\bibitem{IASMSK}{S. Inouye, M. R. Andrews, J. Stenger, H.-J. Miesner, D. M.
Stamper-Kurn, and W. Ketterle}, Nature \textbf{392}, 151 (1998)
\bibitem{KSFBCCCS}{L. Khaykovich, F. Screck, G. Ferrari, T. Bourdel, J.
Cubizolles, L. D. Carr, Y. Castin, and C. Salomon}, Science \textbf{296},
1290 (2002)
\bibitem{LDTRZMLDIM}{E. Lucioni, B. Deissler, L. Tanzi, G. Roati, M. Zaccanti,
M. Modugno, M. Larcher, F. Dalfovo, M. Ignuscio, and G. Modugno}, Phys. Rev.
Lett. \textbf{106}, 230403 (2011)
\bibitem{BAA}{D. M. Basko, I. L. Aleiner, and B. L. Altshuler}, Annals of
Physics \textbf{321}, 1126 (2006)
\bibitem{DS}{D. L. Shepelyansky}, Phys. Rev. Lett. \textbf{70}, 1787 (1993)
\bibitem{LBKSF} {T. V. Laptyeva, J. D. Bodyfelt, D. O. Krimer, Ch. Skokos, and
S. Flach}, EPL \textbf{91}, 30001 (2010) 
\bibitem{MAP}{M. Mulansky, K. Ahnert, and A. Pikovsky}, Phys. Rev. E
\textbf{83}, 026205 (2011)
\bibitem{LDM}{M. Larcher, F. Dalfovo, and M. Modugno}, Phys. Rev. A
\textbf{80}, 053606 (2009)
\bibitem{GS}{I. Garc\'{i}a-Mata and D. Shepelyansky}, Phys. Rev. E \textbf{79},
026205 (2009)
\bibitem{GBF}{G. Gligori\'{c}, J. D. Bodyfelt, and S. Flach}, EPL
\textbf{96}, 30004 (2011)
\bibitem{FKS}{S. Flach, D. O. Krimer, and Ch. Skokos}, Phys. Rev. Lett.
\textbf{102}, 024101 (2009)
\bibitem{LBF}{T. V. Laptyeva, J. D. Bodyfelt, and S. Flach}, EPL
\textbf{98}, 60002 (2012)
\bibitem{MP}{M. Mulansky, and A. Pikovsky}, arxiv:1207.072v1 (2012)
\bibitem{SF}{S. Flach}, Chem. Phys. \textbf{375}, 548 (2010)
\bibitem{KKFA} {G. Kopidakis, S. Komineas, S. Flach, and S. Aubry}, Phys. Rev.
Lett. \textbf{100}, 084103 (2008)
\bibitem{VF}{R. A. Rodrigo, and S. Flach}, Phys. Rev. E \textbf{79}, 016217
(2009)
\bibitem{SKKF}{Ch. Skokos, D. O. Krimer, S. Komineas, and S. Flach}, Phys. Rev.
E \textbf{79}, 056211 (2009)
\bibitem{FW}{S. Flach and C. R. Willis}, Phys. Rep. \textbf{295}, 181 (1998)
\bibitem{FI}{F. M. Izrailev}, Phys. Rep. \textbf{196}, 299 (1990)
\bibitem{CD}{S. W. Cleveland, and S. J. Devlin}, J. Am. Stat. Assoc.
\textbf{83}, 596 (1988)
\bibitem{Ho}{J. Hoffman}, \textit{Numerical Methods for Engineers and
Scientists} (McGraw-Hill, New York) (1992)
\bibitem{DF}{E. Doron, and S. Fishman}, Phys. Rev. Lett. \textbf{60}, 867 (1988)
\end{thebibliography}
\end{document}